# Anomaly Classification in Distribution Networks Using a Quotient Gradient System

Hamid Khodabandehlou, Iman Niazazari, Hanif Livani, and M. Sami Fadali

*Abstract*— **The classification of anomalies or sudden changes in power networks versus normal abrupt changes or switching actions is essential to take appropriate maintenance actions that guarantee the quality of power delivery. This issue has increased in importance and has become more complicated with the proliferation of volatile resources that introduce variability, uncertainty, and intermittency in circuit behavior that can be observed as variations in voltage and current phasors. This makes diagnostics applications more challenging. This paper proposes using quotient gradient system (QGS) to train two-stage partially recurrent neural network to improve anomaly classification rate in power distribution networks using high-fidelity data from micro-phasor measurement units (µPMUs). QGS is a systematic approach to finding solutions of constraint satisfaction problems. We transform the µPMUs data from the power network into a constraint satisfaction problem and use QGS to train a neural network by solving the resulting optimization problem. Simulation results show that the proposed supervised classification method can reliably distinguish between different anomalies in power distribution networks. Comparison with other neural network classifiers shows that QGS trained networks provide significantly better classification. Sensitivity analysis is performed concerning the number of µPMUs, reporting rates, noise level and early versus late data stream fusion frameworks.**

## I. INTRODUCTION

### A. Motivation

Due to the frequent disruptive events in power networks, developing a data-driven event diagnostics framework for maintaining the regular operation of the system is of paramount importance. Establishing such a framework, not only assists system operators in extracting useful information such as the cause or location of events, but also aids in other applications, such as preventive maintenance. Preventive maintenance is an effective action regarding time, cost, safety, avoiding unexpected outages, and maintenance crew utilization.

Although disruptive events may not cause immediate equipment failure, they gradually lead to permanent failure. Hence, a comprehensive study of disruptive event classification in power systems is beneficial and will eventually lead to increasing the life expectancy of critical assets. With the expansion in the number of high-fidelity metering devices (i.e., phasor measurement units (PMUs) and micro-PMUs (µPMUs)) in power systems, data-driven diagnostics frameworks have become feasible for utilities and system operators.

### B. Related Works

Power quality disturbance classification using the wavelet transform has been extensively investigated in the literature [1]-[4]. In [5], the authors used the wavelet transform and support vector machine (SVM) for feature representation and classification of power quality events and disturbances. Masoum et al. proposed a novel approach for detection and classification of disturbances in power systems [6]. The distorted signal is first denoised using the discrete wavelet transform (DWT), and the dominant features are then fed to a wavelet network classifier. A wavelet-based neural network method for detection and classification of power quality disturbances in power systems was proposed in [7]. In general, DWT-based methods are an effective technique for dimensionality reduction. They reduce the computational time while preserving the accuracy. The multi-resolution analysis used in DWT is a beneficial tool for achieving quicker data mining and reducing data storage. However, compared to other methods, DWT based methods do not perform well in the presence of noise.

There are several other methods for the classification of power quality disturbance. Several authors have studied the detection and classification of power quality disturbances using SVM [8]-[11]. In [12], two different classification methods are compared. The first method represents features using principal component analysis (PCA). The representation is the input to a multi-class SVM. In the second method, feature representation and classification use the autoencoders and softmax classifiers, respectively. Although the SVM can handle high dimensional data well, it requires a proper choice of the kernel function and suffers from overfitting. In addition, their computational complexity increases drastically with the size of the training set.

Fuzzy-based methods for the classification of power quality disturbances were investigated in [13]-[16]. Manikandan et al. proposed a new method based on a sparse signal decomposition algorithm on hybrid dictionaries for detection and classification of power quality disturbances [17]. Fuzzy methods are particularly useful for pre and post data analysis. However, their drawback is the high resource consumption involved [18].

Anomaly classification is a significant task that has been extensively investigated within various research areas in the literature. Anomaly detection and classification using the S-

This paper is based upon work supported by the National Science Foundation under Grant No. IIA-1301726.

H. Khodabandehlou, I. Niazazari, H. Livani and M. Sami. Fadali are with the Department of Electrical and Biomedical Engineering, University of Nevada, Reno, NV, 89557 USA (e-mail: hkhodabandehlou@nevada.unr.edu, niazazari@nevada.unr.edu, hlivani@ieee.org, fadali@unr.edu).

transform have been widely studied in [19]-[22]. The event classification based on S-transform and probabilistic neural network (PNN) needs fewer features compared to the wavelet based method and outperforms the feedforward multilayer (FFML) and learning vector quantization (LVQ) methods [19]. Unlike the wavelet transform, the features obtained from the S-transform have physical significance and can quantify the disturbances. The time-frequency resolution of S-transform makes it a good candidate for event classification. In addition, it results in better accuracy in the presence of noise. Using the modular neural network yields better accuracy and less training time, compared to a single NN [22].

In [23], a new approach for event classification and localization in power system was proposed based on the hyperbolic S-transform (HS) and radial basis function neural network (RBFNN). The HS-transform was applied to the input signal to generate the correspondent time-frequency contours, phase contours, and absolute phase components. The extracted numerical indices then are fed to RBFNN for classification. In [24], the authors presented a method for the power quality disturbance classification using the S-transform and based on genetic algorithm (GA) and PNN. The dominant features of data captured by the S-transform are fed to the PNN for the automatic classification of disturbances. Finally, GA is used to optimize the smoothing parameters of the PNN and improve the overall classification accuracy. Although this method has some advantages, the performance of the genetic algorithm depends on the proper choice of mutation and crossover methods as well as the proper choice of initial population and has a very slow convergence rate. Hence, the PNN's are very slow to train and need much more memory than multilayer perceptron networks.

In [25], an autoencoder based neural network is proposed for the classication of the abnormal events in the distribution system,A wide variety of methods are available to train neural networks for anomaly classification, such as the scaled complex conjugate algorithm [21], improved generalized adaptive resonance theory [26], Marquardt Levenberg [27] and learning vector quantization combined with genetic algorithm [28]. Nevertheless, there are promising constrained optimization approaches in the mathematics literature that have not been used to train neural networks for classification applications.

In [30],[31], it is shown that an optimization approach, known as the quotient gradient method, offers many advantages in training neural networks over conventional approaches. This stems from the fact that QGS finds the global minimum of the squared error criterion optimized in neural network training rather than the local minima to which other training approaches often converge. The method is a trajectory-based methodology that uses trajectories of a nonlinear dynamical system, the quotient gradient system (QGS), to find feasible solutions of constrained optimization problems [30]. The trajectories of the QGS converge to its equilibrium point, which is also the solutions to the optimization problem.

The quotient gradient method is a systematic approach to find the feasible solutions of the constraint satisfaction problems. It transforms the constraint satisfaction problem into an unconstrained minimization problem that defines the QGS. The equilibrium points of the QGS are local minima of the unconstrained minimization problem as well as the feasible solutions of the constraint satisfaction problem. In [31], the authors used the QGS to train a single stage fully recurrent neural network for nonlinear system identification and compared the results with those of error backpropagation.

QGS does not have user dependent variables, such as learning rate. Unlike Newton-based methods, it does not require a huge number of measurements. Furthermore, QGS does not need proper choice of starting point and does not require a considerable amount of memory. The independence from initial values, not having user dependent variables along with reasonable training time, makes QGS an interesting alternative training approach for neural networks. In this paper, we propose the use of QGS to train a two-layer partially recurrent neural network for anomaly classification and localization in power distribution networks and compare the results with the results of error backpropagation and the results of genetic algorithm trained networks.

*C. Contribution*

This paper proposes a novel PMU-data-driven framework for classification and localization of disruptive events in distribution grids. Four different events are considered in this paper: (1) malfunctioned capacitor bank switching; (2) malfunctioned regulator on-load tap changer (OLTC) switching; (3) grid reconfiguration; and (4) normal abrupt load change. The proposed classification algorithm is developed using neural networks trained with the quotient gradient method. The novelties of this method are as follows:

- It proposes a unified anomaly classification and localization in distribution grids using PMU data. The proposed method distinguishes the type and the location of abnormal events in a single framework that simplifies the existing data-driven methods by combining the abnormal event classification and localization steps.
- The proposed classifier is quite resilient in the presence of measurements noise and can distinguish classes better than genetic algorithm and backpropagation based trained neural networks. In the presence of large measurement noise, the QGS trained network outperforms GA trained network and error backpropagation trained network by a significant margin.
- The proposed method is based on a two-layer neural network that improves the classification accuracy compared to state-of-the-art NN-based classifiers. The first layer distinguishes the malfunctioned capacitor bank switching and reconfiguration events. The second layer distinguishes between events that have similar signatures and are much more difficult to separate, malfunctioned OLTC and abrupt load changing events.

The rest of this paper is organized as follows: The problem statement is presented in Section II. Section III and IV present the QGS method and its application to neural networks training, respectively. Simulation results are presented in Section V. Finally, the conclusion is presented in Section VI.

II. PROBLEM STATEMENT

Disruptive events occur intermittently in power systems interrupting normal operation. Finding a mechanism to detect, classify, and localize the source and location of these events prevents further damage to equipment and power outages.

The schematic diagram of the proposed event diagnostics framework is presented in Fig. 1. High-resolution metering devices, such as PMUs, are installed at several nodes in distribution systems. The measured data are then transmitted to the data storage and archiving center through communication links, such as LTE networks [32]. In the final stage, post-event processing, including events classification and localization, is carried out using the received PMU data.

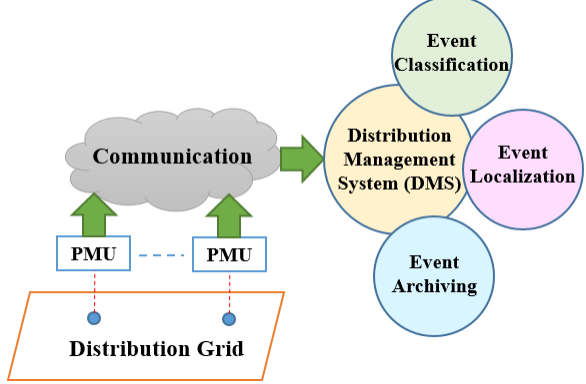

Fig. 1. PMU data-driven event classification in distribution systems

Four different events are studied in this paper: (1) malfunctioned capacitor bank switching; (2) malfunctioned regulator OLTC switching; (3) grid reconfiguration; and (4) normal abrupt load change. The first two classes are disruptive events, and the last two are normal events. Malfunctioned capacitor bank switching events occur because of failure in the mechanical switches of transformers. It takes about one cycle, i.e., 16.67 *msec,* for a capacitor bank to switch [33]. In a malfunctioned regulator OLTC switching, the tap is dislocated and then relocated to its original position. Aging and degradation of the selector switches may cause malfunctioned regulator OLTC switching. The on-load tap changer switching takes about 30-200 *msec* [34].

The third event is network reconfiguration in which one recloser opens, and another one closes. Opening and closing distribution reclosers take about five cycles, i.e., 83 *msec* [35]. The last class is the abrupt load changing occurs due to an increase or decrease in demand in some nodes of the network. In this paper, events are first categorized based on their types and then based on their locations. Therefore, each event with a specific type and at a particular location is assigned to a class.

PMUs measure voltage magnitudes (pu), voltage angles (degree), current magnitudes (pu), and current angles (degree). For classification, we calculate: (i) the difference between two consecutive PMU samples, i.e., the change in voltage magnitude between two successive samples ($|v(n+1)|-|v(n)|$), (ii) the change in the voltage angle between two consecutive samples ($\delta_v(n+1)-\delta_v(n)$), (iii) the change in the current magnitude between two successive samples ($|i(n+1)|-|i(n)|$), and (iv) the change in the current angle between two consecutive samples ($\delta_i(n+1)-\delta_i(n)$). It is assumed that the sequence of events, i.e., the pre- and the during-event sequence is identified before initiating the classification process with an algorithm such as [36]. Next, the feature matrix is constructed using the current magnitude (pu), and current angle (degree) of the pre-event and during-event PMU samples along with the difference between the successive pre-event and during-event samples, as shown in Fig. 2. The feature matrix is the input to the neural network for classification.

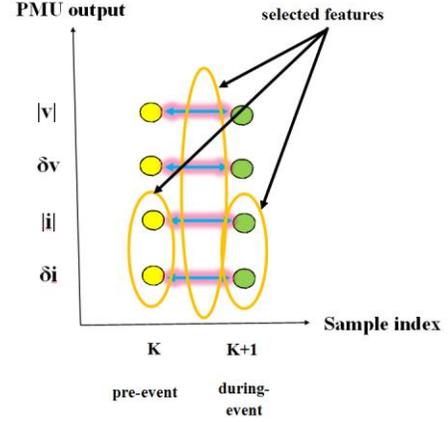

Fig. 2. Feature selection process

### III. QUOTIENT GRADIENT SYSTEM

A system of linear or nonlinearly constrained equations appears in many fields, including engineering and science. Lee and Chiang used the trajectories of a nonlinear dynamical system, the QGS, to find feasible solutions of the constraint satisfaction problem [29]. This section reviews the work of Lee and Chiang. Consider the following constraint satisfaction problem

$$C_I(y) < 0 \quad (1)$$
$$C_E(y) = 0, y \in R^{n-l}$$

where $C_I(y)$ is the set of inequality constraints and $C_E(y)$ is the set of equality constraints and $y$ is the vector of unknown variables to be found. To guarantee the existence of the solution, $C_I = (c_1, \ldots, c_l)^T: R^{n-l} \to R^l$ and $C_E = (c_{l+1}, \ldots, c_m)^T: R^{n-l} \to R^{m-l}$ need to be smooth. The constraint satisfaction problem can be rewritten as the unconstrained minimization problem

$$\min_x f(x) = \frac{1}{2}\|h(x)\|^2, \quad x = (y, s) \in R^n \quad (2)$$

$$h(x) = \begin{bmatrix} C_I(y) + \hat{s}^2 \\ C_E(y) \end{bmatrix} \in R^m, \hat{s}^2 = (s_1^2, \ldots, s_l^2)^T \quad (3)$$

where $\hat{s}$ is set of introduced slack variables needed to write the inequality constraints as equality constraints. Local minima of (2) are feasible solutions of (1). The QGS is a nonlinear dynamical system based on the constraints given by

$$\dot{x} = F(x) = -\nabla f(x) \coloneqq -D_x h(x)^T h(x) \quad (4)$$

The stable equilibrium points of the QGS are local minima of the unconstrained minimization problem, which are the possible feasible solutions of the constraint satisfaction problem [29].

Note that the stable equilibrium points of the GQS are not necessarily in the feasible region of the constraint satisfaction problem. In such cases, the QGS needs to escape from the stability region of that stable equilibrium point and enter the stability region of another stable equilibrium point. This process must be repeated until the QGS enters the stability region of a stable equilibrium point or until it reaches the stopping criterion. The QGS can reach the stability region of the stable equilibrium point by integrating from an initial point, which can be infeasible. The QGS can escape from the stability region by integrating backward in time. The eigenvalues of the Jacobian matrix of the QGS are a measure of stability or instability [29],[30].

## IV. APPLICATION OF QGS TO NEURAL NETWORKS

Neural networks have been successfully used in many engineering applications, including system identification and pattern recognition among the others. In this study, we use a three-layer recurrent neural network with one hidden layer. Fig. 3 depicts the internal structure of the neural network. $\mathbf{u}(k)$, $\mathbf{z}(k)$ and $\hat{\mathbf{y}}(k)$ are the input, internal state and output vectors of the network and are respectively defined as

$$\begin{aligned}\mathbf{u}(k) &= [u_1(k) \quad \dots \quad u_n(k)]^T \\ \mathbf{z}(k) &= [z_1(k) \quad \dots \quad z_m(k)]^T \\ \hat{\mathbf{y}}(k) &= [\hat{y}_1(k) \quad \dots \quad \hat{y}_q(k)]^T\end{aligned} \quad (5)$$

The governing equation of the network is

$$\mathbf{z}(k) = \boldsymbol{\psi}\big(W\mathbf{u}(k) + P\mathbf{z}(k-1)\big)$$
$$\hat{\mathbf{y}}(k) = V\mathbf{z}(k) \quad (6)$$

where $\boldsymbol{\psi}$ is the activation function of the hidden layer nodes. The activation function is the tangent hyperbolic function

$$\psi(x) = \tanh(x) = \frac{e^x - e^{-x}}{e^x + e^{-x}} \quad (7)$$

For a network with $n$ inputs, $m$ hidden layer nodes and $q$ outputs, $W$ and $V$ are $m \times n$ and $q \times m$ matrices respectively and $P$ is, a $m \times m$ diagonal matrix. The cost function for training network s the Sum of Squared Errors (SSE)

$$SSE = \sum_{k=1}^{N} \mathbf{e}(k)^T \mathbf{e}(k)$$
$$= \sum_{k=1}^{N} \big(\hat{\mathbf{y}}(k) - \mathbf{y}(k)\big)^T \big(\hat{\mathbf{y}}(k) - \mathbf{y}(k)\big) \quad (8)$$

where $\mathbf{e}(k)$ is the error between network output, $\hat{\mathbf{y}}(k)$, the target output $\mathbf{y}(k)$, and $N$ is the number of training samples.

QGS is used to minimize the SSE to find the optimal values of the network weights. QGS provides a systematic method to find the local minima of the unconstrained minimization problem of (2). To train the neural network using the QGS, we write the training set as the equality constraints of (1). The resulting minimization problem is to minimize the sum of the squared errors and is solved using the QGS. The QGS finds the set of local minima of the optimization problem and the local minimum with the lowest cost is the global minimum of the minimization problem. If $N$ measurement samples are available, the constraints are

$$\mathbf{h}(\mathbf{x}) = [h_i(\mathbf{x})], i = 1,2,\dots,N$$
$$h_i(\mathbf{x}) = V\boldsymbol{\psi}\big(W\mathbf{u}(i) + P\mathbf{z}(i-1)\big) - \mathbf{y}(i) \quad (9)$$

where $\mathbf{x}$ is the vector of network parameters comprising all the elements of $V$ and $W$ and nonzero elements of $P$ with

$$V = \begin{bmatrix}\mathbf{v}_1^T \\ \vdots \\ \mathbf{v}_m^T\end{bmatrix}_{q \times m} \quad W = \begin{bmatrix}\mathbf{w}_1^T \\ \vdots \\ \mathbf{w}_m^T\end{bmatrix}_{m \times n} \quad P = diag(\mathbf{p})_{m \times m} \quad (10)$$

In (10), $diag(\mathbf{p})$ denotes a diagonal matrix with the elements of $\mathbf{p}$ as its diagonal elements. Using (10), $\mathbf{x}$ can be expressed as

$$\mathbf{x} = [x_i]_{n_p \times 1} = [\mathbf{v}_1, \dots, \mathbf{v}_m, \mathbf{w}_1, \dots, \mathbf{w}_m, \mathbf{p}]^T \quad (11)$$

$$n_p = m \times (n + q + 1)$$

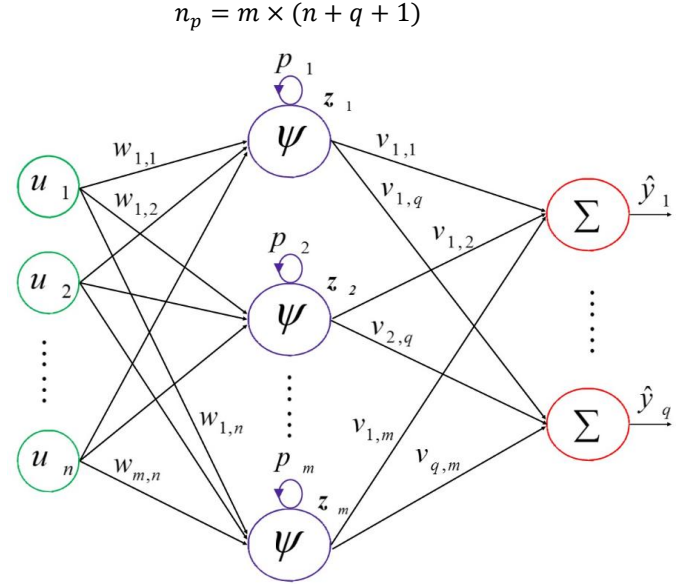

Fig. 3. Structure of the neural network

In addition to the constraints of (1), upper and lower bounds are imposed on the network parameters. The upper and lower bounds are inequality constraints that can be transformed into equality constraints by introducing slack variables. The QGS for training neural network is written as

$$\dot{\mathbf{x}} = -\mathbf{f}(\mathbf{x}) = -D_x \mathbf{h}(\mathbf{x})^T \mathbf{h}(\mathbf{x}) \quad (12)$$

where

$$D_x \mathbf{h}(\mathbf{x}) = \begin{bmatrix}\dfrac{\partial h_1(\mathbf{x})}{\partial \mathbf{x}} \\ \vdots \\ \dfrac{\partial h_N(\mathbf{x})}{\partial \mathbf{x}}\end{bmatrix}_{N \times n_p} \quad (13)$$

Thus, the problem of training neural network reduces to a series of forward and backward integration of the QGS to find the local minima of the SSE minimization in (8). After finding the set of local minima, the one with the lowest cost is chosen as the global minimum of the optimization problem [30].

## V. SIMULATION RESULTS

To validate the performance of the QGS-trained neural network, we use it to classify anomalies in the IEEE 123-bus test systems.

The modified IEEE 123-bus system is shown in Fig.4. The network is composed of (1) four three-phase capacitor banks at buses 51, 57, 83, and 108, (2) four voltage regulators at bus 149-150, bus 9-14, bus 25-26, and bus 67-160, (3) 91 loads at different buses, (4) six normally closed and six normally opened reclosers at different buses shown in Fig. 4.

The line model used for this system is a "Pi" model with shunt capacitance. The line parameters, R, X, and C matrices are properly chosen for the unbalanced system. The loads in the network are defined with their nominal active and reactive powers. The loads in the simulation are modeled with one of the following three implementations
1) Constant P and constant Q,
2) Constant Z (or constant impedance)

3) Constant I (or constant current magnitude)

The voltage regulator has a control, which can change the line drop compensator setting by adjusting R, X, primary of CT, and PT ratio. There are 33 different tap positions on the regulator that allows for -10% to 10% variation from the nominal value.

There are five PMUs located on five buses, 1, 13, 18, 60, and 97 for streaming data. The PMU located at bus 1 serves as the angle reference for other PMUs. The other four PMUs measure voltage at bus 13, 18, 60, and 97 respectively, and current from bus 152 to 13, bus 135 to 13, bus 160 to 60, and bus 197 to 97, respectively. It is assumed that PMUs only stream the steady-state signals and transient states are ignored. Therefore, all the simulations are performed in the steady-state mode using OpenDSS, a comprehensive electrical power system simulation tool primarily developed for distribution systems [38]. The PMUs used in this paper have two reporting rates: (1) 60 sample per second (SPS), such as SEL 651 [39]; and (2) 120 SPS, such as µPMUs developed at the University of California, Berkeley [40].

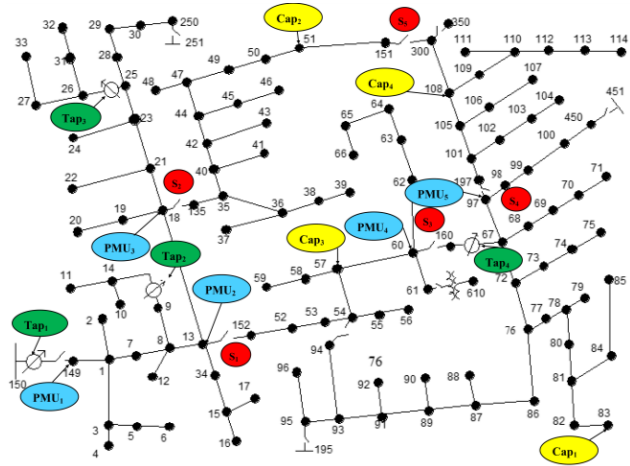

Fig. 4. The modified IEEE 123-bus system

Fig. 5 shows the PMU voltage magnitude of phase *a* at bus 60 over one second, 60 samples, corresponding to four different events as (a) a malfunctioned capacitor bank switching at bus 57, (b) a malfunctioned OLTC switching of the voltage regulator between bus 149 and 150, (c) an abrupt load change at bus 71, and (d) a reconfiguration as the result of switching actions of $S_2$ and $S_5$. In the malfunctioned capacitor switching, the capacitor switches off and then switches back on. The switching changes the amount of reactive power in the system and, subsequently, causes voltage variations in the capacitor bank substation and neighboring substations. In the malfunctioned OLTC switching, the tap changer dislocates to an unwanted position and then returns to its original position. Shifting the position of the tap changer causes a change in the resistance and reactance of the regulator and changes the turn ratio of the regulator and the Y-bus of the system. In the reconfiguration event, opening one switch and closing another changes the system topology and the system Y-bus. This changes the voltage and current in the system.

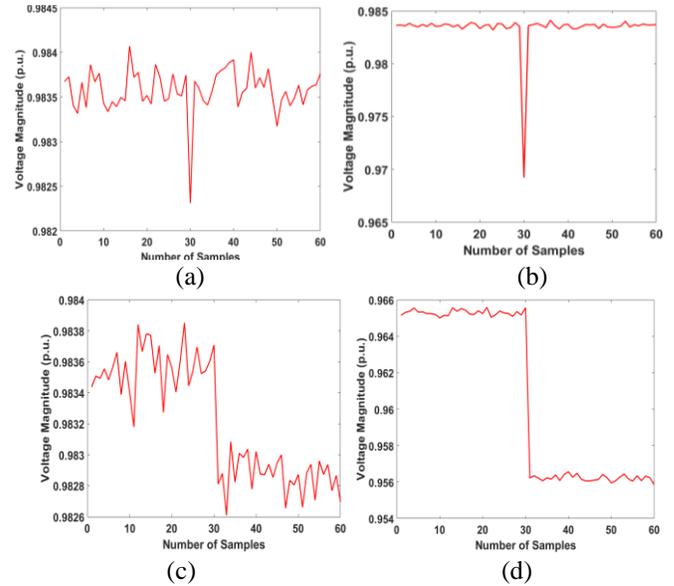

Fig. 5. PMU voltage magnitude of phase *a* at bus 60 over one second a) malfunctioned capacitor bank switching, b) malfunctioned OLTC switching c) abrupt load changing c) reconfiguration

Data-driven event diagnostics methods rely on the availability of historical datasets with event information. However, the number of abnormal events with known class type is limited in practice. Therefore, three main solutions are used to overcome this challenge [41]. (1) All unknown events are treated as a single class and a supervised event classification is carried out based on available field data. (2) An unsupervised clustering algorithm is used to identify all the possible events, and a supervised classification is then utilized based on the field data. (3) All the disruptive and normal events are simulated using the state-of-the-art simulation packages and the trained classifiers are tested based on the new datasets. Because of lack of sufficient field data for disruptive and normal abrupt change events, this paper adopts the latter solution.

To classify the anomalies according to their types and locations, events are divided into 13 different classes where each class is labeled with a tag indicating its types and location. For example, if an event is classified as class 2, it is concluded that a malfunctioned capacitor switching occurred at bus 51, as shown in Fig. 4. If an event is classified as class 7, it is concluded that a malfunctioned OLTC switching occurred between buses 25 and 26. This procedure is followed for other classes to label all the classes.

To create enough experiments for malfunctioned three-phase capacitor bank switching events (classes 1 to 4), it is assumed that the capacitor switching occurs at different loading conditions. In this network, there are 91 different loads, and for each of the loads, ten different loading levels ranging from 50% to 140% of the average level with 10% increment are simulated. Therefore, an overall of 910 different experiments is simulated for each capacitor bank switching class. The same number of experiments are simulated for malfunctioned OLTC switching events (classes 5 to 8). For a normal abrupt load change (class 9), it is assumed that only one load can suddenly change at a time. The abrupt load change is simulated based on 5%, 10%, 15%, 20%, and 25% increase or decrease in its power. Therefore, the overall number of experiments in class 9 is 910. Finally, for the reconfiguration events, it is assumed that one recloser opens and another one closes. Four different reconfiguration events are simulated, 1) $S_1$ opens & $S_5$ closes; 2) $S_2$ opens & $S_5$ closes; 3) $S_3$ opens & $S_5$ closes; and 4) $S_4$ opens & $S_5$ closes. An overall of 910 experiments is simulated for the reconfiguration events (classes 10 to 13). The classification accuracy is calculated as

$$Accuracy = \frac{number\ of\ accurate\ classification}{total\ number\ of\ test\ cases} \quad (14)$$

Preliminary results revealed that distinguishing between classes 6, 7, 8 and 9 (OLTC switching of regulators two, three, four and abrupt load changing) is difficult due to the similarity of their PMU data. Therefore, a two-layer neural network for anomaly classification is proposed and shown in Fig. 6. The extracted feature vector is first fed into the neural network in the first layer. This layer distinguishes the capacitor bank switching and the recloser actions versus the remaining classes. The second layer classifies the OLTC switching versus the abrupt load changes. The optimal number of hidden layer nodes is $m = 8$ for the first network and $m = 6$ for the second neural network. All the parameters of networks are initialized with random values from a zero-mean normal distribution with standard deviation $\sigma = 0.5$. We have 900 samples for each type of anomaly. We randomly choose 100 samples from each type of events to train the neural networks and use the remaining samples as the evaluation dataset. The QGS finds 15 different local minima for the optimization problem. We select the parameter values corresponding to local minimum with the best generalization capability as the optimal parameters of the neural network. The neural network with the optimal parameters is used to classify new events.

Table II shows the confusion matrix where the performance of the trained neural networks is evaluated. Class 6, the OLTC switching of the regulator between buses 9 and 14 is correctly classified with 88% accuracy and 10% misclassified as class 7 and 2% misclassified as class 5. Class 7, the OLTC switching of the regulator between buses 25 and 26 is classified with 84% accuracy, while 16% of the events are misclassified as class 6. The classification accuracy of class 8 is 98% with 2% misclassification as class 9. The classification accuracy of class 9 is 94% with 6% misclassification as class 8. The classification accuracies for the remaining classes of 100% validates the acceptable performance of the proposed two-layer neural network for event classification in distribution grids. Note that the overall classification accuracy is 96%.

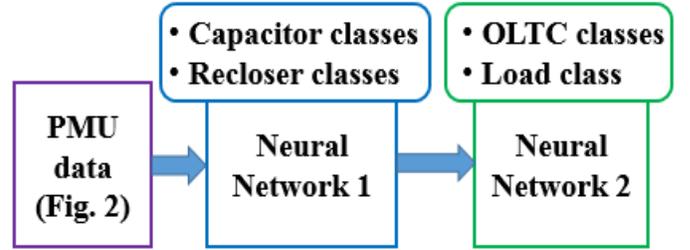

Fig. 6. Two-layer NN-based events classification

Table II. Classification confusion matrix

| | | Output Class | | | | | | | | | | | | |
|---|---|---|---|---|---|---|---|---|---|---|---|---|---|---|
| | | 1 | 2 | 3 | 4 | 5 | 6 | 7 | 8 | 9 | 10 | 11 | 12 | 13 |
| Target Class | 1 | 100% | 0% | 0% | 0% | 0% | 0% | 0% | 0% | 0% | 0% | 0% | 0% | 0% |
| | 2 | 0% | 100% | 0% | 0% | 0% | 0% | 0% | 0% | 0% | 0% | 0% | 0% | 0% |
| | 3 | 0% | 0% | 100% | 0% | 0% | 0% | 0% | 0% | 0% | 0% | 0% | 0% | 0% |
| | 4 | 0% | 0% | 0% | 100% | 0% | 0% | 0% | 0% | 0% | 0% | 0% | 0% | 0% |
| | 5 | 0% | 0% | 0% | 0% | 100% | 0% | 0% | 0% | 0% | 0% | 0% | 0% | 0% |
| | 6 | 0% | 0% | 0% | 0% | 2% | 88% | 10% | 0% | 0% | 0% | 0% | 0% | 0% |
| | 7 | 0% | 0% | 0% | 0% | 0% | 16% | 84% | 0% | 0% | 0% | 0% | 0% | 0% |
| | 8 | 0% | 0% | 0% | 0% | 0% | 0% | 0% | 98% | 2% | 0% | 0% | 0% | 0% |
| | 9 | 0% | 0% | 0% | 0% | 0% | 0% | 0% | 6% | 94% | 0% | 0% | 0% | 0% |
| | 10 | 0% | 0% | 0% | 0% | 0% | 0% | 0% | 0% | 0% | 100% | 0% | 0% | 0% |
| | 11 | 0% | 0% | 0% | 0% | 0% | 0% | 0% | 0% | 0% | 0% | 100% | 0% | 0% |
| | 12 | 0% | 0% | 0% | 0% | 0% | 0% | 0% | 0% | 0% | 0% | 0% | 100% | 0% |
| | 13 | 0% | 0% | 0% | 0% | 0% | 0% | 0% | 0% | 0% | 0% | 0% | 0% | 100% |

*a)*      *PMU Reporting Rate Analysis*

To assess the performance of the proposed method with respect to the reporting rate of PMU, the number of hidden layer nodes of the first and second neural networks must first be determined. In this scenario, the number of hidden layer nodes is 10 for the first neural network and 4 for the second. Increasing the PMU reporting rate does not have a significant impact on the classification rate but improves the overall classification accuracy by 0.66%. This is because a 60 SPS reporting rate is fast enough to capture the events in the grid. Table III shows the overall classification accuracies for 60 and 120 SPS reporting rates.

Table III. Overall classification accuracies for 60 and 120 SPS

| PMU reporting rate | 60 sps | 120 sps |
|---|---|---|
| Accuracy | 96% | 96.66% |

*b)*      *Measurement Noise Analysis*

PMU data may include errors or measurement noise. To analyze the effect of measurement noise, Gaussian white noise is added to the PMU data stream. The measurement noises with standard deviations of $\sigma^2 = 0.005$, $\sigma^2 = 0.01$, $\sigma^2 = 0.02$, and $\sigma^2 = 0.05$ of the reported phasor values are added to the PMU data, and the noisy data is then used to train and evaluate the neural networks. Table IV shows the overall classification accuracies with noisy data. As it can be seen from Table IV, increasing the noise level, decreases the classification accuracy.

Table IV. Overall classification accuracies with different level of noise

| $\sigma^2$ | 0.005 | 0.01 | 0.02 | 0.05 |
|---|---|---|---|---|
| Accuracy | 95.33% | 93.11% | 90.22% | 83.77% |

*c)  Number of PMUs*

To analyze the impacts of number of installed PMUs on the performnace of the proposed method, the neural networks are trained using the data stream from 3 PMUs (PMU 2, 3, 4), 2 PMUs (PMU 3, 4), and only PMU 4, respectively. Table V shows the overall classification accuracies for different numbers of PMUs. The classification accuracy increases with the number of PMUs. Fewer PMUs result in fewer features from the grid and decreases the classification accuracy. However, with even one PMU the proposed method achieves a limited accuracy level, which shows the practicality of the proposed data-driven method, as utilities install PMUs over the years.

Table V. Classification accuracy with different number of installed PMUs

| Number of PMUs | 1 | 2 | 3 | 4 |
|---|---|---|---|---|
| Accuracy | 64.66% | 83.11% | 92.22% | 96% |

*d)  Boosting Scenario*

In the boosting scenario, we add the misclassified data to the training set and retrain the neural networks. After training with the new data set, a new set of test data is fed to the neural network, the misclassified data is added to training data, and the network is trained again. This process is repeated three times and then we test the network with a new test data set. Boosting does not appear to have a significant effect on the anomaly classification results and only improves the overall accuracy by a modest 0.44%. Table VI shows the overall classification accuracy with the boosting scenario.

Table VI. Classification accuracy with boosting scenario

| Training scenario | Normal | Boosting |
|---|---|---|
| Accuracy | 96% | 96.44% |

*e)  Comparison with Traditional Neural Networks*

To show the effectiveness of the QGS in training neural networks for event classification and localization, we compare the overall classification accuracy with those obtained by neural networks trained using the classical error backpropagation (EBP) and the genetic algorithm (GA). The genetic algorithm training uses the MATLAB optimization toolbox. An initial population size of 10000, with top fitness scaling, Roulette selection, adaptive feasible mutation and scattered crossover resulted in the best performance for the GA-trained neural network.

Table VII shows the results of the three different training methods. The overall classification accuracies are calculated with measurement noise variance $\sigma^2 = 0.05$. The QGS and GA result in superior performance compared to backpropagation. This is expected since both methods are global optimization approaches. Although, the QGS- and GA-trained network errors are within the acceptable range, the QGS-trained network has better generalization capability and outperforms the GA-trained network by 6.5%.

Table VII. Classification accuracy comparison with GA- and EBP-based NN

| NN Training method | QGS | GA | EBP |
|---|---|---|---|
| Accuracy | 83.77% | 77.33% | 70.44% |

## VI. CONCLUSION

This paper proposes the use of the quotient gradient system (QGS) to train recurrent neural networks (NNs) for classifying and localizing anomalies in power distribution networks. The proposed algorithm is developed to distinguish two classes of anomalies, malfunctioned capacitor bank switching and malfunctioned on-load tap changer (OLTC) switching versus normal recloser switching and abrupt load changing. To enhance the accuracy of state-of-the-art events classifiers, NNs reformulate a constraint satisfaction problem as an unconstrained minimization problem to be solved using the QGS approach. The input data are the phasor measurement units (PMUs) data from the grid. The performance of the proposed algorithm is evaluated using the simulation results in the IEEE 123-bus system. The sensitivity analysis with respect to the reporting rate and number of installed PMUs, noise level, boosting scenario, and the comparison with genetic algorithm- and error backpropagation-based NNs validate the performance of the proposed anomaly classification and localization method.

REFERENCES

[1] W. M. Lin, C. H. Wu, C. H. Lin, and F .S. Cheng, "Detection and classification of multiple power-quality disturbances with wavelet multiclass SVM," *IEEE Transactions on Power Delivery*, vol. 23, no. 4, pp. 2575-2582, 2008.
[2] S. Santoso, E.J. Powers, W. M. Grady, and P. Hofmann, "Power quality assessment via wavelet transform analysis," *IEEE Transactions on Power Delivery*, vol. 11, no. 2, pp. 924-930, 1996.
[3] A. M. Gaouda, M. M. A. Salama, M. R. Sultan, and A. Y. Chikhani, "Power quality detection and classification using wavelet-multiresolution signal decomposition," *IEEE Transactions on Power Delivery*, vol. 14, no. 4, pp.1469-1476, 1999.
[4] M. Khoshdeli, I. Niazazari, R. J. Hamidi, H. Livani, and B. Parvin, "Electromagnetic Transient Events (EMTE) Classification in Transmission Grids," In *Proc. Power and Energy Society General Meeting* (PESGM), pp. 1-5,2017.
[5] Z. Moravej, A. A. Abdoos, and M. Pazoki, "Detection and classification of power quality disturbances using wavelet transform and support vector machines," *Electric Power Components and Systems*, vol. 38, no. 2, pp. 182-196, 2009.
[6] M. A. S. Masoum, S. Jamali, and N. Ghaffarzadeh,"Detection and classification of power quality disturbances using discrete wavelet transform and wavelet networks," *IET Science, Measurement & Technology*, vol. 4, no. 4, pp. 193-205, 2010.
[7] Z. L. Gaing, "Wavelet-based neural network for power disturbance recognition and classification," *IEEE Transactions on Power Delivery*, vol. 19, no. 4, pp.1560-1568, 2004.
[8] P. Janik and T. Lobos, "Automated classification of power-quality disturbances using SVM and RBF networks," *IEEE Transactions on Power Delivery*, vol. 21, no. 3, pp.1663-1669, 2006.
[9] H. Erişti and Y. Demir, "A new algorithm for automatic classification of power quality events based on wavelet transform and SVM," *Expert systems with applications*, vol. 37, no. 6, pp. 4094-4102, 2010.
[10] P. G. Axelberg, I. Y. H. Gu, and M. H. Bollen, "Support vector machine for classification of voltage disturbances," *IEEE Transactions on Power Delivery*, vol. 22, no. 3, pp. 1297-1303, 2007.
[11] X. F. Song and J. C. Chen, "Classification method of dynamic power quality disturbances based on SVM," *Electric Power Automation Equipment*, vol. 26, no. 4, pp. 39-42, 2006.
[12] I. Niazazari and H. Livani, "Disruptive Event Classification using PMU Data in Distribution Networks," In *Proc. Power and Energy Society General Meeting* (PESGM), pp. 1-5, 2017

[13] B. K. Panigrahi and V. R. Pandi, "Optimal feature selection for classification of power quality disturbances using wavelet packet-based fuzzy k-nearest neighbour algorithm," *IET Generation, Transmission & Distribution*, vol. 3, no. 3, pp. 296-306, 2009.
[14] M. V. Chilukuri and P. K. Dash, "Multiresolution S-transform-based fuzzy recognition system for power quality events," *IEEE Transactions on Power Delivery*, vol. 19, no. 1, pp. 323-330, 2004.
[15] G. S. Hu, J. Xie, and F. F. Zhu, "Classification of power quality disturbances using wavelet and fuzzy support vector machines," In *Proc. of International Conference on Machine Learning and Cybernetics*, vol. 7, pp. 3981-3984, Aug. 2005.
[16] J. Huang, M. Negnevitsky, and D. T. Nguyen, "A neural-fuzzy classifier for recognition of power quality disturbances," *IEEE Transactions on Power Delivery*, vol. 17, no. 2, pp. 609-616, 2002.
[17] M. S. Manikandan, S. R. Samantaray, and I. Kamwa, "Detection and classification of power quality disturbances using sparse signal decomposition on hybrid dictionaries," *IEEE Transactions on Instrumentation and Measurement*, vol. 64, no. 1, pp. 27-38, 2015.
[18] H. Kaur, G. Singh, and J. Minhas, "A review of machine learning based anomaly detection techniques," *International Journal of Computer Applications Technology and Research*, vol. 2, no. 2, pp. 185-187, 2013.
[19] S. Mishra, C. N. Bhende, and B. K. Panigrahi, "Detection and classification of power quality disturbances using S-transform and probabilistic neural network," *IEEE Transactions on Power Delivery*, vol. 23, no. 1, pp. 280-7, Jan 2008.
[20] F. Zhao and R. Yang, "Power-quality disturbance recognition using S-transform," *IEEE Transactions on Power Delivery*, vol. 22, no. 2, pp. 944-950, Apr. 2007.
[21] R. Kumar, B. Singh, D. T. Shahani, A. Chandra, and K. Al-Haddad, "Recognition of power-quality disturbances using S-transform-based ANN classifier and rule-based decision tree," *IEEE Transactions on Industry Applications*, vol.51, no. 2, pp. 1249-1258, Mar./Apr. 2015.
[22] C. N. Bhende, S. Mishra, and B. K. Panigrahi, "Detection and classification of power quality disturbances using S-transform and modular neural network," *Electric Power Systems Research*, vol. 78, no. 1, pp. 122-128, 2008.
[23] S. R. Samantaray, P. K. Dash, and G. Panda, "Power system events classification using pattern recognition approach," *International Journal of Emerging Electric Power Systems*, vol. 6, no.1, 2006.
[24] K. Manimala and K. Selvi,"Power Disturbances Classification Using S-Transform Based GA–PNN," *Journal of The Institution of Engineers* (India), vol. 96, no.3, pp. 283-295, 2015.
[25] I. Niazazari, and H. Livani,"A PMU-data-driven disruptive event classification in distribution systems," *Electric Power Systems Research*, vol. 157, pp. 251-260, 2018.
[26] K. S. Yap, C. P. Lim, M. T. Au, "Improved GART neural network model for pattern classification and rule extraction with application to power systems," *IEEE Transaction on Neural Netw.*, vol. 22, no. 12, pp. 2310-2323, Dec. 2011.
[27] D. Biswas, P. M. Adhikari, A. De, "An artificial neural network based power swing classification technique," *India Conference* (INDICON), 2014 Annual IEEE, 11-13 Dec. 2014
[28] O. Sen, S. Zhengxiang, W. Jinhua, C. Degui, "Application of LVQ neural networks combined with genetic algorithm in power quality signals classification," *International Conference on Power System Technology*, pp. 491-495, 2002.
[29] J. Lee, H. D. Chiang, "Quotient gradient methods for solving constraint satisfaction problems," In *Proc. IEEE Int. Symp. Circuits and Systems*, May 2001
[30] H. Khodabandehlou and M. Sami Fadali, "A Quotient Gradient Method to Train Artificial Neural Networks," *In Proc. Int. Joint Conf. Neural Networks (IJCNN), Anchorage, USA*, 2017
[31] H. Khodabandehlou and M. Sami Fadali, "Training Recurrent Networks as a Constraint Satisfaction Problem," *arXiv:1803.07200v6 [cs.LG]*
[32] M. Elattar and J. Jasperneite, "Using LTE as an access network for internet-based cyber-physical systems," *11th IEEE World Conference on Factory Communication Systems* (WFCS), pp. 1-7, May 2015.
[33] L.P.Hayes,(2005). First Energy Corp, Akron, OH, [Online] Available: http://castlepowersolutions.biz/Components/Joslyn%20Hi-Voltage/Technical%20Papers/First_Energy_Corporation_Paper_Doble_Conference_2005_VBM_Failures.pdf
[34] R. Jongen, E. Gulski, K. Siodła, J. Parciak and J. Erbrink, "Diagnosis of degradation effects of on-load tap changer in power transformers," *ICHVE International Conference on High Voltage Engineering and Application*, Poznan, pp. 1-4, 2014.
[35] Siemens Vacuum Recloser 3AD. [Available online] https://w3.siemens.com/powerdistribution/global/SiteCollectionDocuments/en/mv/outdoor-devices/catalogue-vaccum-recloser-3AD_en.pdf
[36] Y. Zhou, R. Arghandeh, I. Konstantakopoulos, S. Abdullah, A.V. Meier and C.J. Spanos, "Abnormal event detection with high resolution micro-PMU data," *Power Systems Computation Conference* (PSCC), pp. 1-7, jun. 2016.
[37] H.Th. Jongen ,P. Jonker and F. Twilt, Nonlinear Optimization in Finite Dimensions: Morse Theory, Chebyshev Approximation, Transversality, Flows, Parametric Aspects. Kluwer Academic, Dordrecht, 2000.
[38] EPRI, "Simulation Tool – OpenDSS", [online] Available http://smartgrid.epri.com/SimulationTool.aspx.
[39] SEL 651 relay. [Online] Available: https://selinc.com/
[40] A. V. Meier, D. Culler, A. McEachern, and R. Arghandeh. "Micro-synchrophasors for distribution systems," In. *Proc*. *Innovative Smart Grid Technologies Conference* (ISGT), pp. 1-5, 2014.
[41] S. Brahma, R. Kavasseri, H. Cao, N.R. Chaudhuri, T. Alexopoulos, and Y. Cui, "Real-Time Identification of Dynamic Events in Power Systems Using PMU Data, and Potential Applications—Models, Promises, and Challenges," *IEEE Transactions on Power Delivery*, vol. 32, no.1, pp. 294-301, 2017.